\documentclass[conference]{IEEEtran}
\IEEEoverridecommandlockouts
\usepackage{cite}
\usepackage{amsmath,amssymb,amsfonts}
\usepackage{algorithmic}
\usepackage{graphicx}
\usepackage{multirow}
\usepackage{multicol}
\usepackage{textcomp}
\usepackage{xcolor}
\def\BibTeX{{\rm B\kern-.05em{\sc i\kern-.025em b}\kern-.08em
    T\kern-.1667em\lower.7ex\hbox{E}\kern-.125emX}}
    
\usepackage{algorithm}  
\usepackage{amsmath}

\usepackage{cleveref}
\crefname{equation}{}{}

\usepackage{subfig}

\usepackage{caption}
\makeatother
\begin{document}

\title{Minimizing Delay in Network Function Visualization with Quantum Computing\\
}

\author{\IEEEauthorblockN{Wenlu Xuan$^{1}$, Zhongqi Zhao$^{1}$, Lei Fan$^{2}$ and Zhu Han$^{1}$}
\IEEEauthorblockA{\textit{Dept. of Electrical and Computer Engineering$^{1}$ and Dept. of Engineering Technology$^{2}$} \\
\textit{University of Houston}}
}

\maketitle

\begin{abstract}
Network function virtualization (NFV) is a crucial technology for the 5G network development because it can improve the flexibility of employing hardware and reduce the construction of base stations. There are vast service chains in NFV to meet users' requests, which are composed of a sequence of network functions. These virtual network functions (VNFs) are implemented in virtual machines by software and virtual environment. How to deploy VMs to process VNFs of the service chains as soon as possible when users’ requests are received is very challenging to solve by traditional algorithms on a large scale. Compared with traditional algorithms, quantum computing has better computational performance because of quantum parallelism. We build an integer linear programming model of the VNF scheduling problem with the objective of minimizing delays, and transfer it into the quadratic unconstrained binary optimization (QUBO) model. Our proposed heuristic algorithm employs a quantum annealer to solve the model. Finally, we evaluate the computational results and explore the feasibility of leveraging quantum computing to solve the VNFs scheduling problem.
\end{abstract}


\section{Introduction}
In recent years, network function virtualization (NFV) has attracted more attention in the revolution of wireless network technology because it can reduce the cost of deploying hardware and improve network flexibility. For traditional network function technology, specific hardware can only process particular functions, which raises many challenges for machine manufacturing and maintenance, and causes waste while adjusting network functions. For NFV technology, virtual network functions (VNFs) are implemented at universal standard servers, which solves the issues mentioned above. The standard server integrates many types of equipment. Every virtual machine (VM) is allocated on standard commercial servers. Each VM realizes one or more VNFs of the network by software or virtual environments \cite{han2015,Mijumbi2016Network}. VNFs, which will be split, can adopt in one more VM with parallel running to reduce processing delay. Briefly speaking, the implementation of different VNFs is based on various software instead of specific hardware. Thus, NFV reduces the difficulty of hardware configuration and ameliorates the compatibility of a network. For the realization of NFV, physical machines (PMs) like standard servers are virtualized as one or more VMs, which attaches much importance to how to allocate VMs at PMs under the consideration of efficiency and costs. This kind of problem is called the VM embedding problem \cite{zhang2019}. Another important problem is how to deploy VMs to process VNFs when users’ request is received. In this paper, we propose an integer linear programming (ILP) model with delay minimization as the objective function to solve the VNFs scheduling problem.

In the NFV network system, VMs are usually located at data centers, and VMs interconnect with others by virtual links. Each VM is configurated with specific computing resources. If the user's request needs to process oversized data packages, the processing delay at VMs cannot be ignored. The transmission delay between two VMs, especially in the same data center, can be neglected because the transmission rate is high. For most users and network service providers, the total delay is a crucial aspect of evaluating the performance of a network, which is identified in the service level agreement. A request of users can be realized by the cooperation of several VNFs, and we introduce a service chain to correspond to the request in the NFV-enabled network \cite{Medhat2017}. A service chain is a sequence of ordered VNFs. The network processes data via VNFs, definitely following the order in the service chain. Consequently, the total delay of processing the request is the whole period that data goes through the corresponding service chain. In this paper, our model aims to optimize the total processing delay of all service chains in the network.

In the real world, the network always receives significant numbers of requests, and it needs to complete the process as soon as possible, and so the desperate need for network operators is an efficient solver that can optimize large-scale resource allocation. Traditional algorithms cannot meet such a requirement, and so we pin the hope on quantum computing. The superiority of quantum computing is based on quantum parallelism, which means a quantum computer searches for possible outcomes simultaneously. As a result, the computation speed of quantum computers is much faster than that of classical computers. It has been proved the speedup of quantum computing in solving certain problems\cite{roland_quantum_2002,childs_exponential_2003,hen_period_2014,somma_quantum_2012}. 

Quantum computing takes advantage of quantum properties, such as quantum superposition and quantum entanglement. The first milestone in quantum computation is the no-cloning theorem, which shows the impossibility of copying an unknown quantum state \cite{wootters1982}. This is one of the fundaments of quantum computing and quantum information. In 1980, Benioff brought up a method of simulating quantum systems by Turing machines \cite{benioff1980}. In 1982, Feynman introduced the conception of quantum computation, and following brought up the idea that universal quantum computing can be realized through quantum systems following quantum mechanics \cite{feynman1982}. In 1985, David Deutsch first proposed a computing paradigm based on quantum mechanics, which is the framework of the modern quantum computer \cite{deutsch1985Quantum}. Many powerful quantum algorithms were proposed by prominent scientists in the past several decades, such as the Deutch-Jozsa Algorithm \cite{Deutsch1992Rapid}, Grover's Algorithm \cite{Grover1996}, Shor's Algorithm \cite{shor1997}, and Quantum Approximate Optimization Algorithm \cite{farhi2014}. These algorithms demonstrate the great potential of quantum computing in many fields.

In recent years, tech giants, like IBM, Google, Microsoft, and D-wave, took the lead in developing quantum computers. D-wave company employs quantum annealing techniques to construct the quantum annealer. Quantum annealing assists a quantum system in reaching the lowest energy state. Compared with other quantum computing models, e.g. the analog quantum model and universal quantum gate model, quantum annealing technique provides more quantum bits in current industry practice, which means the quantum annealing hardware has more powerful computational performance. The quantum annealer can efficiently solve the quadratic unconstrained binary optimization (QUBO) problem by using the Ising model, which describes the energy state with coupling qubits interaction and externally applied fields \cite{ising1925}. Therefore, the QUBO model can leverage a qubit system via embedding methods to get an optimal solution. It has been proved that many combinatorial optimization problems can be rewritten in the QUBO form, and it facilitates the application of quantum annealing machines \cite{glover2019QuantumBridge}.

This paper formulates the VNFs scheduling problem as an ILP model with the optimization of delay, which is not easy to be solved by classical algorithms. We transfer the ILP model into the QUBO form and propose a heuristic algorithm to solve it using the D-Wave hybrid solver. We study several cases with different parameters and different scales to evaluate the performance of the D-Wave hybrid solver in solving our model. Our key contributions are as follows:

\noindent \vspace{-0.5cm}
\begin{itemize}
\item We propose an ILP model for the NVFs scheduling problem, and then we reformulate the model as the QUBO model, which can be solved by the quantum annealing machine.
\item We propose a heuristic algorithm to quickly find a feasible solution, which can help strengthen our QUBO model. We also demonstrate the efficiency of our algorithm in multiple experiments.
\item We employ quantum computing to solve the VNFs scheduling problem. Our work shows the possibility of using quantum computing to allocate resources in NFV.
\end{itemize}

The rest of this paper is organized as follows. Section \mbox{II} introduces related work about VNF scheduling problems and quantum computing applications. Section \mbox{III} illustrates the NFV system. In Section \mbox{IV}, we describe the ILP model of the VNFs scheduling problem and reformulates the ILP model as the QUBO model, and we propose a heuristic algorithm to embed the QUBO onto quantum annealing hardware. Section \mbox{V} shows the case study results of this problem using quantum computing. Finally, Section \mbox{VI} is the conclusion of the whole paper.

\section{Related Work}

\subsection{VNFs Scheduling Problem}

Because of the importance of the user request processing time, many researchers study the optimal processing delays in the VNFs scheduling problem \cite{diez2019,gouareb2018,qu2016Network}. In \cite{diez2019}, Diez \textit{et al}. implemented NFV into cloud radio access networks, essential in 5G. They took split selection and scheduling into consideration while minimizing traffic delay globally and partially. The results showed that partial optimization is close to the exact optimal solution. Geared \textit{et al}. \cite{gouareb2018} formulated a complex VNFs chaining and placement model considering queuing delay in virtual links and edge clouds. They also analyzed different queuing models in the same situation. Their proposed schemes satisfied the stringent quality of service and meet service-level agreement requirements for both horizontal scaling and vertical scaling. In \cite{qu2016Network}, researchers studied a VNFs scheduling problem with minimizing the total delay, including processing delay and transmission delay. They also considered the dynamic virtual link bandwidth while formulating the model. A genetic-algorithm-based method was proposed to get the optimal solution due to the high complexity of the model. 

To improve the practicality of proposed models, many researchers tend to study VNFs scheduling problems from multiple aspects including minimizing delay, maximizing throughput, optimizing cost, and improving reliability \cite{yang2019,luizelli2015,ren2020,qu2017A}. \cite{yang2019} presented an integer nonlinear programming model to illustrate the VNFs scheduling problem. They developed a heuristic algorithm to achieve the minimum delay in different scenarios. The resiliency of NFV was also studied by constructing a more reliable virtual network. Luizelli \textit{et al}. \cite{luizelli2015} leveraged the ILP method to formulate the VNF placement and chaining problem. The ILP was designed to minimize delays while guaranteeing the utilization efficiency of resources. They also proposed a heuristic algorithm to solve it and compared this method with other optimal approaches in different scenarios. Their algorithm found a better solution with under the consideration of end-to-end delays. In \cite{ren2020}, Ren \textit{et al}. presented a delay-sensitive NFV-enabled multicasting problem in mobile edge clouds. They aimed to minimize the implementation cost of the request and to maximize the system throughput. An approximation algorithm was proposed to solve the model without delay requirement, and a heuristic algorithm was developed to solve the complex model. \cite{qu2017A} presented a reliability-aware and delay constrained optimization model in NFV-enabled networks. The system model deployed backup VNFs over multiple paths to improve the reliability of the virtualization network. The paper proposed a  mixed integer linear programming (MILP) model to jointly optimize reliability, end-to-end delays, and resource consumption. A heuristic algorithm based on greedy-$k$-shortest paths was used to solve the MILP model. The results showed that this algorithm had better performance than other schemes in finding optimal solutions.    

\subsection{Quantum Computing}

Benefited from the technology development of controlling quantum particles and constructing quantum hardware, quantum computation has attracted more attention in recent years. Some scientists attempted to use quantum computation to solve optimization problems in wireless networks \cite{kim2019Leveraging,kim2020Towards,alanis2014Quantum-Assisted,alanis2015Non-Dominated}. Researchers \cite{kim2019Leveraging} employed quantum computing to solve sizeable multiple-input multiple-output (MIMO) problems in centralized radio networks. They analyzed the performance under different modulations, which demonstrated that quantum computing is generally valuable in these cases. In the binary phase shift keying (BPSK) communication system, quantum annealers can assist the network in serving 48 users with an extremely low bit error rate. \cite{kim2020Towards} investigated the boundary between classical and quantum computing in wireless systems. Hybrid classical-quantum computing methods was based on current quantum computers, which also called noisy intermediate-scale quantum devices. They evaluated the performance of reverse annealing techniques in the hybrid classical-quantum frames and compared it with that of forward annealing and also novel forward-reverse annealing. Alanis \textit{et al}. \cite{alanis2014Quantum-Assisted} developed a non-dominated quantum optimization (NDQO) algorithm for multi-objective routing problems. NDQO illustrated an approximate optimal performance compared with the state-of-the-art evolutionary algorithms. However, they found that the NDQO algorithm was infeasible to search for an available solution with the number of nodes increasing and the number of routes increasing exponentially. Consequently, they proposed a non-dominated quantum iterative optimization (NDQIO) algorithm in \cite{alanis2015Non-Dominated}. The NDIQO algorithm run on the quantum hardware parallelization framework. It had a good performance of serving routing in wireless multihop networks despite the complexity reduction compared with the NDQO algorithm.  

Some researchers also tried to use quantum annealers to solve classical NP-hard problems \cite{cruz2019,venturelli2016}. In \cite{cruz2019}, the minimum multicut (MMC) problems were transformed into QUBO formulations using two different methods. They studied a particular case of the MMC problem on the family of random connected trees. The QUBO model of this case was processed at the D-Wave machine to get an optimal solution. Venturelli \textit{et al}. \cite{venturelli2016} built a simple model for the job-shop scheduling problem with the makespan minimization. They formulated this model in the QUBO form and embed it on D-Wave chips. Some strategies of fine-tuning parameters and graph-embedding were also presented in this paper. Their results showed that pre-processing using classical algorithms is efficient in this situation. 

\section{System Model}

\begin{figure}[tbp]
	\centering
	\subfloat[]{\label{fig1a}\includegraphics[scale=0.52]{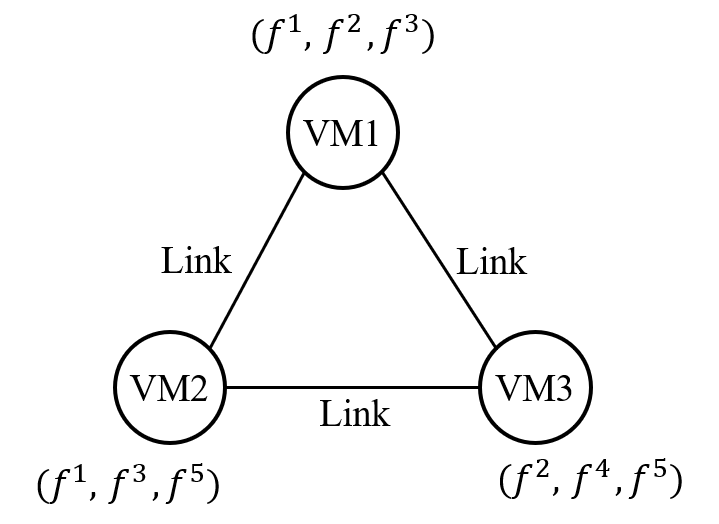}}\quad
	\subfloat[]{\label{fig1b}\includegraphics[scale=0.35]{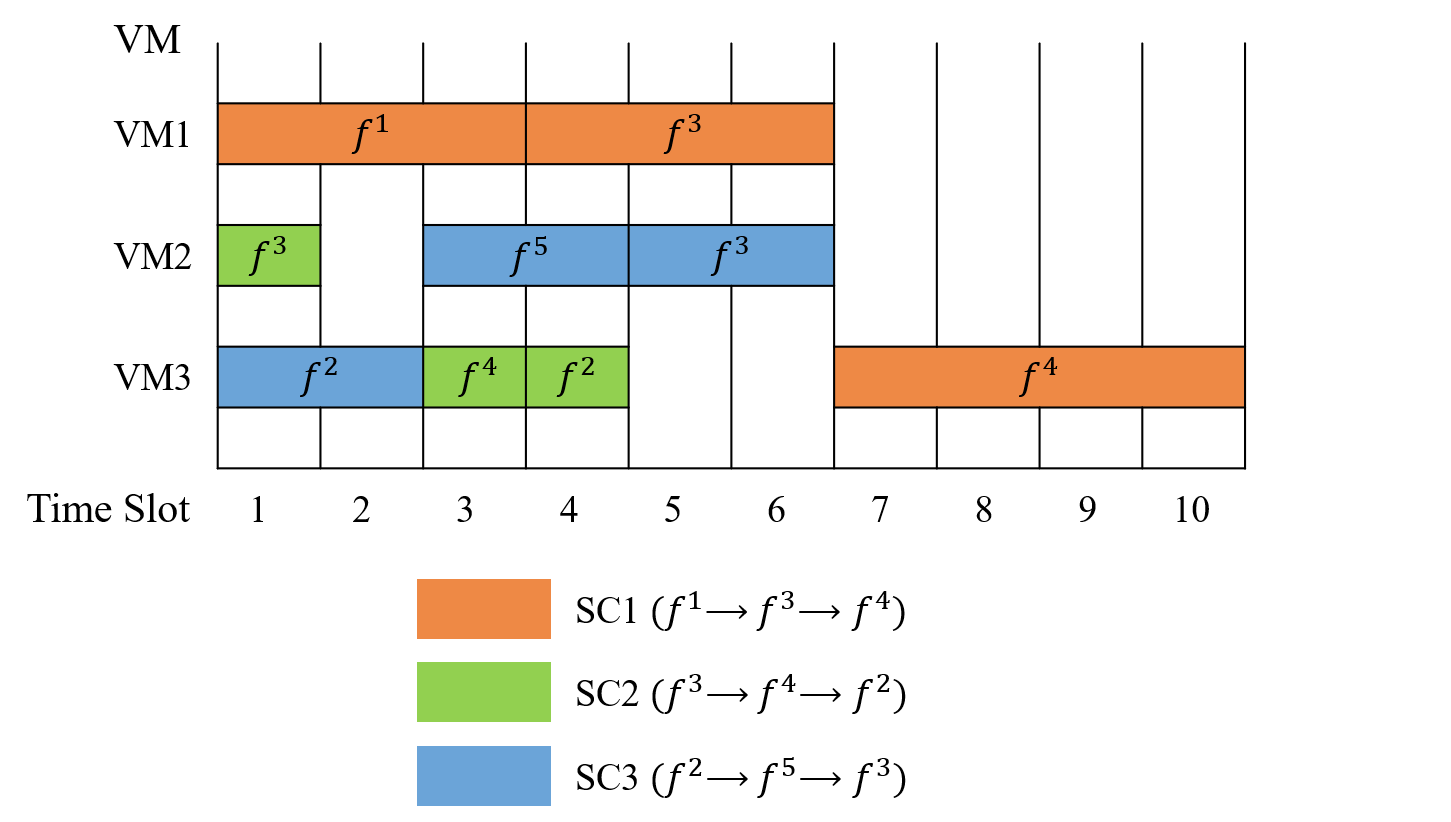}}\\	
	\caption{Example of NFV system $(a)$ a simple NFV system $(b)$ a possible arrangement of service chains.}
	\label{fig1}
\end{figure}

In our system model, the network offers $K$ types of VNFs to satisfy user's requirements. $F =\left\{f^{1}, f^{2},...,f^{k},..., f^{K} \right\}$ denotes the set of VNFs and $f^{k}$ denotes the $k^{th}$ type of functions. Any VM can serve one or more than one function, and any function $f^{k}$ can be configured on any VM. VMs can be divided into several groups, and for each group, these VMs serve the same VNFs and have the same computing capacity. According to the users' requirement, these functions compose a service chain $i$ to processing data. Thus, we distinguish different VNF instances of the same network functions by denoting them as $f^{k}_{ij}$, which means the $j^{th}$ function in service chain $i$ belongs to the $k^{th}$ type of functions. We assume that each VM can serve one function at a time. There is only one virtual link between any two VMs in the network, and we denote the virtual link between VM $m$ to VM $n$ as $l_{(m,n)}$. In our NFV network, all VMs are embedded in PMs located in data centers, and the transmission between any two VMs is high-speed, and so the transmission delay can be neglected. We only consider the minimum total processing delay to scheduling VNFs of service chains.

Workload $W_{ij}$, corresponding to the data package required processing, will be processed on VM $m$, $m\in V^{k}_{ij}$. The computing capability of VM $m$ is denoted by $C_{m}$. Thus, the processing time $t_{ijm}$ is given by $W_{ij}/C_{m}$. We set the system model as a discrete-time model, and so the controller's working time will be divided into several time slots with the length of $\Delta T$.  $T_{ijm}$ is the minimum integer that is equal to or larger than $(t_{ijm}/ \Delta T)$. It can be said that the number of time slots occupied by $t_{ijm}$ is $T_{ijm}$.  

Fig. \ref{fig1} is an example of how the controller schedules the VMs to satisfy the request requirements from the users. Suppose that there are three virtual machines, VM1, VM2, and VM3. VM1 can process the functions $f^{1}$, $f^{2}$, and $f^{3}$. VM2 can process the functions $f^{1}$, $f^{3}$, and $f^{5}$. VM3 can process the functions $f^{2}$, $f^{4}$, and $f^{5}$. According to the requests from users, the network receives service chains, SC1 ($f^{1}\longrightarrow f^{3}\longrightarrow f^{4}$) with $4 MB$, SC2 ($f^{3} \longrightarrow f^{4} \longrightarrow f^{2}$) with $0.8 MB$, and SC3 ($f^{2} \longrightarrow f^{5} \longrightarrow f^{3}$) with $2 MB$. In this example, the data size of packages won't change after processed by any VNFs. Every time slot has a length of $1s$. We assume that VM1 can process $1.5 MB$ per second, and the computation rate of VM1 is $1.5$ times that of VM2 and VM3. The controller arranges VM1 to process $f^{1}$ of SC1, and the processing delay is 3 time slots. In SC1, the second function that needs to be processed is $f^{3}$. Fortunately, VM1 can also process $f^{3}$, and so the controller arranges VM1 to process $f^{3}$ because VM1 processes data faster than the other two VMs. The processing delay of $f^{3}$ is also 3 time slots. After that, the results of $f^{3}$ are transferred to VM3 via the link between VM1 and VM3, and then, the processing delay of $f^{4}$ is 4 time slots at VM3. The total delay of SC1 is 10 time slots. The first function of SC2, $f^{3}$, is processed at VM2, which costs 1 time slot. The only VM that can process $f^{4}$ is VM3, so the results of $f^{3}$ are transmitted to VM3. However, VM3 is occupied by SC3, so SC2 needs to wait until VM3 finishes processing $f^{2}$ of SC3. After 1 time slot, VM3 starts to process $f^{4}$ of SC2, and the processing time is $0.8 s$, which means 1 time slot is occupied. After that, the controller arranges VM3 to process $f^{2}$ of SC2, and the process occupies 1 time slot. Even though $f^{4}$ and $f^{2}$ of SC2 are processed at VM3 successively, the whole process is not continuous.  At the end of the time slot of serving $f^{4}$ of SC2, the controller will determine if the processing finishes or not, and then arrange other functions to be processed at VM3. The total delay of SC2 is 4 time slots. The first function of SC3 is $f^{2}$, and it is processed at VM3, which costs 2 time slots.  After that, the results are sent to VM2 through the link between VM2 and VM3. The second function of SC3 is $f^{5}$, which is processed at VM2 for 2 time slots, and then the third function $f^{3}$ is also processed for 2 time slots at VM2. The total delay of SC3 is 6 time slots. The processing delay of all three service chains is 20 time slots, which means $20 s$. Finally, the controller will evaluate this arrangement and determine whether the total delay of all service chains arrives at the minimum. If not, the controller will rearrange VMs to process these functions.

\begin{table*}[htbp]
\renewcommand\arraystretch{1.5}
\caption{List of Notations}

\label{table1}
\begin{center}

\begin{tabular}{|c|p{15cm}|}
\hline
\textbf{Notation}& \textbf{Description}\\
\hline
$i, i'$& index of service chain; $i,i'\in \left\{1,2,\cdots,I\right\}$ \\
\hline
$j, j'$& index of the sequence of function in service chain; $j,j'\in \left\{1,2,\cdots,J\right\}$ \\
\hline
$f^{k}$& the $k^{th}$ type of functions,  $ k\in\left\{1,2,\cdots,K\right\}$ \\
\hline
$F$& the set of all $f^{k}$ \\
\hline
$f^{k}_{ij}$& the $j^{th}$ function in service $i$ belongs to the $k^{th}$ type of functions  \\
\hline
$f^{k'}_{i(j+1)}$& the $(j+1)^{th}$ function in service $i$ belongs to the $(k')^{th}$ type of functions; $k'\in\left\{1,2,\cdots,K\right\}$; $ j\in\left\{1,2,\cdots,(J-1)\right\}$ \\
\hline
 $m, m'$& index of VM; $m\in\left\{1,2,\cdots,M\right\}$ \\
\hline
 $n$& index of VM; $n\in\left\{1,2,\cdots,M\right\}$ \\
\hline
$V^{k}_{ij}$& the set of VMs which can serve $f^{k}_{ij}$\\
\hline
$V^{k'}_{i(j+1)}$& the set of VMs which can serve $f^{k'}_{i(j+1)}$\\
\hline
$l_{(m,n)}$& the virtual link between VM $m$ and VM $n$;\\
\hline
$\Delta T$& the length of each time slot;\\
\hline
$t$&the $t^{th}$ time slot; $t\in\left\{1,2,\cdots,T_{max}\right\}$ \\
\hline
$W_{ij}$& the workload of processing $f^{k}_{ij}$ \\
\hline
$C_{ijm}$& the computing capability of VM $m$ which can serve $f^{k}_{ij}$\\
\hline
$t_{ijm}$& the time length of processing $f^{k}_{ij}$ on VM $m$\\
\hline
$T_{ijm}$& the number of time slots occupied by processing $f^{k}_{ij}$ on VM $m$\\
\hline
$s_{iJ}$& the finish time of processing the last function of service chain $i$\\
\hline
$x_{ijm}$& equals to 1, if VM $m$ is used to process $f^{k}_{ij}$; otherwise, equals to 0\\
\hline
$y_{ijmt}$& equals to 1, if VM $m$ is used to process $f^{k}_{ij}$ in the time slot $t$; otherwise, equals to 0\\
\hline
$z_{ijmt}$& equals to 1, if VM $m$ starts to process $f^{k}_{ij}$ at the beginning of the time slot $t$; otherwise, equals to 0\\
\hline
$p_{ijmt}$& equals to 1, if VM $m$ finishes processing $f^{k}_{ij}$ at the beginning of the time slot $t$; otherwise, equals to 0\\

\hline
\end{tabular}
\label{tab1}
\end{center}
\noindent \vspace{-0.5cm}
\end{table*}

\section{Problem Formulation and Algorithm}

\subsection{ILP Formulation}

We develop an ILP model to describe the NVF scheduling problem. This model aims to minimize the total delay of all service chains in the network, and we use the finish time of the last function as the total delay of the corresponding service chain. All notations used in the model and their descriptions are listed in Table \ref{table1}. For the ILP model, the objective function, constraints, and their explanations are listed below.
\noindent \vspace{-0.5cm}

\noindent\begin{center}
\begin{equation}
\min_{s_{iJ}}\quad\tilde{s}=\sum\limits^{I}_{i=1}  s_{iJ}\label{eq1}
\end{equation}

\begin{equation}
s_{iJ} = \sum\limits_{m=1}^{M}\sum\limits_{t=1}^{T_{max}} p_{iJmt}\cdot (t-1)\cdot \Delta T, \quad\forall i.\label{eq2}
\end{equation}
\end{center}
Eq. \eqref{eq1} is the objective function. Eq. \eqref{eq2} shows how to calculate the finish time of any service chain. If $p_{1J34}$ is equal to 1, it means that the service chain $1$ finishes to be processed at the beginning of the $4^{th}$ time slot on VM $3$, and so the processing delay of service chain 1 is 3 time slots. 

In addition, we have the following constraints.
\noindent \vspace{-0.4cm}
\noindent\begin{center}

\begin{equation}
\sum\limits_{m\in V^{k}_{ij}} x_{ijm}  = 1 , \quad \forall i,j.\label{eq3}
\end{equation}
\noindent \vspace{-0.1cm}
\begin{equation}
x_{ijm}  = \sum\limits_{t=1}^{T_{max}}z_{ijmt} ,\quad \forall i,j,m.\label{eq4}
\end{equation}
\par\end{center}
Constraint \eqref{eq3} indicates that any function $f^{k}_{ij}$ can be processed on only one VM. Notice that in \eqref{eq3} only the VM in the set $V_{ij}^{k}$ can be selected. Constraint \eqref{eq4} shows the relationship between $x_{ijm}$ and $z_{ijmt}$. If and only if $f^{k}_{ij}$ is allocated to VM $m$, this VM can start processing $f^{k}_{ij}$ at some point.

\noindent \vspace{-0.8cm}
\noindent\begin{center}

\begin{equation}
\sum\limits_{i=1}^{I}\sum\limits_{j=1}^{J} y_{ijmt} \leq 1, \quad\forall m,t.\label{eq5}
\end{equation}
\par\end{center}
Constraint \eqref{eq5} shows that each VM can process at most one function in one time slot. For example, if $y_{1234} = 1$, which means that VM $3$ processes the second function of service chain $1$ in the $4^{th}$ time slot, VM $3$ cannot processes other functions in this time slot.

\noindent\begin{center}
\noindent \vspace{-0.5cm}
\begin{equation}
y_{ijmt} \leq x_{ijm}, \quad\forall i,j,m,t.\label{eq6}
\end{equation}

\par\end{center}
Constraint \eqref{eq6} indicates the relationship between $x_{ijm}$ and $y_{ijmt}$. If at time $t$, VM $m$ need to handle function $f_{i,j}^k$, which means $y_{ijmt} = 1$, then  $x_{ijm} = 1$.

\noindent \vspace{-0.8cm}
\noindent\begin{center}

\begin{equation}
\sum\limits_{t=1}^{T_{max}} y_{ijmt} =T_{ijm}\cdot x_{ijm} , \quad\forall i,j;\quad m\in V^{k}_{ij}. \label{eq7}
\end{equation}

\par\end{center}
Constraint \eqref{eq7} ensures that required total time $T_{ijm}$ for processing function $f_{ij}^{k}$ must be satisfied. Notice that in \eqref{eq7} only the VM in the set $V_{ij}^{k}$ can be selected because if and only if VM $m$ can process function $f_{ij}^{k}$, $T_{ijm}$ exists.

\noindent\begin{center}
\noindent \vspace{-0.3cm}
\begin{equation}
 z_{ijmt} + p_{ijmt} \leq 1,\quad\forall i,j,m,t.\label{eq8}
\end{equation}

\par\end{center}

\noindent\begin{center}
\noindent \vspace{-0.4cm}
\begin{equation}
y_{ijm(t-1)} - y_{ijmt} + z_{ijmt} - p_{ijmt} =0,\quad\forall i,j,m,t.\label{eq9}
\end{equation}

\par\end{center}
Constraint \eqref{eq8} makes sure that $z_{ijmt}$ and $p_{ijmt}$ cannot be equal to $1$ at the same time, according to the definition of $z_{ijmt}$ and $p_{ijmt}$. Constraint \eqref{eq9} shows the logical relationship between $y_{ijmt}$, $z_{ijmt}$ and $p_{ijmt}$. For example, suppose that $y_{1111} = 0$ and $y_{1112} = 1$. Eq. \eqref{eq9} constrains that $z_{1112}$ must equal $1$ and $p_{1112}$ must equal $0$. For another example, suppose that $y_{1111} = 1$ and $y_{1112} = 0$. Eq. \eqref{eq9} constrains that $z_{1112}$ must equal $0$ and $p_{1112}$ must equal $1$. 

\noindent \vspace{-0.8cm}
\noindent\begin{center}

\begin{equation}
\sum\limits_{\alpha=1}^{T_{ijm}}z_{ijm(t-\alpha+1)} \leq y_{ijmt} ,\quad \forall i,j,t;\quad m\in V^{k}_{ij}.\label{eq10}
\end{equation}

\begin{equation}
\begin{split}
&\sum\limits_{m\in V_{ij}^{k}}\sum\limits_{\beta=1}^{T_{max}}p_{ijm(t-\beta+1)} \geq z_{i(j+1)m't} ,\\
&\qquad\qquad\forall i,j,t; \quad m'\in V^{k'}_{i(j+1)}.\label{eq11}
\end{split}
\end{equation}
\par\end{center}
Constraint \eqref{eq10} guarantees that once the VM starts processing the function $f^{k}_{ij}$, the VM must process it for required time. 
Constraint \eqref{eq11} means that the next function of the service chain must be processed after the processing of the one before it. 

\noindent \vspace{-1.2cm}
\noindent\begin{center}

\begin{equation}
\begin{split}
 &x_{ijm} = y_{ijmt} = z_{ijmt} = p_{ijmt} = 0,\\
 &\qquad\qquad\forall i,j,t; \quad m\notin V_{ij}^{k}. \label{eq12}
 \end{split}
\end{equation}

\begin{equation}
\begin{split}
 \sum\limits_{m\in V_{ij}^{k}}\sum\limits_{t=1}^{T_{max}}z_{ijmt} =\sum\limits_{m\in V_{ij}^{k}}\sum\limits_{t=1}^{T_{max}} p_{ijmt} = 1,\quad\forall i,j. \label{eq13}
 \end{split}
\end{equation}

\par\end{center}
Constraint \eqref{eq12} shows that $x_{ijm}$, $y_{ijmt}$, $z_{ijmt}$, and $p_{ijmt}$ must be equal to $0$ if the VM cannot process the function $f^{k}_{ij}$. Constraint \eqref{eq12} ensures that for any function $f^{k}_{ij}$, only one $z_{ijmt}$, and one $p_{ijmt}$ can be equal to $1$ because the function $f^{k}_{ij}$ can be only processed for one time.

\subsection{QUBO Formulation}

D-Wave quantum annealers can only solve the optimization problem in the QUBO formulation. To leverage quantum annealers, we need to transform the ILP model into the QUBO formulation. The definition of QUBO is as follows:

\noindent \vspace{-1cm}
\noindent\begin{center}
\begin{equation}
\min_{x} \quad f(x)=x^{T}Qx,\label{eq14}
\end{equation}
\par\end{center}

where $x$ is the vector of binary variables, and $Q$ is an upper-diagonal matrix or symmetric matrix. As the definition shows above, there is only an objective function and no constraints in the QUBO formulation. All constraints in our model must be reformulated into quadratic penalties, and then be added to the original objective function. We choose the value of penalty coefficients according to the influence of original constraints in searching for the optimal solution. The principles of transforming classical constraints as equivalent penalties are listed in Table \ref{table2}, where $x_{1} ,x_{2}$ and $x_{3}$ are binary variables. $r_{l}$ is a binary slack variable. $a_{l}$ and $b$ are constants. $P$ is the penalty coefficient. The transformed results of \cref{eq1,eq2,eq3,eq4,eq5,eq6,eq7,eq8,eq9,eq10,eq11,eq12,eq13} are listed in Appendix \cref{appendices}. 

\begin{table}[htbp]
\renewcommand\arraystretch{2}
\caption{List of Constraint-Penalty Pairs}
\noindent \vspace{-0.4cm}
\label{table2}
\begin{center}

\begin{tabular}{|l|l|}
\hline
\textbf{Constraint}& \textbf{Equivalent Penalty}\\
\hline
$x_{1} + x_{2} = 1 $& $P(x_{1} + x_{2} - 1)^{2} $ \\
\hline
$x_{1} + x_{2} + x_{3} \leq 1 $& $P(x_{1}x_{2} + x_{1}x_{3} + x_{2}x_{3}) $ \\
\hline
$x_{1} + x_{2} \leq x_{3} $ & $P(x_{1} + x_{2} - x_{3} + \sum_{l} a_{l}r_{l} )^{2} $\\
\hline
$x_{1} + x_{2} = b $ & $P(x_{1} + x_{2} - b)^{2} $ \\
\hline
\end{tabular}
\end{center}
\end{table}

\subsection{Proposed Algorithm}

We propose a heuristic algorithm to employ the D-Wave solver to solve our model. Due to the limitation of qubits on the D-Wave QPU server, cases with too many variables cannot be solved. On the one hand, to let the solver handles as many variables as possible, we turn to the D-Wave hybrid solver, which employs classical computation to assist quantum annealing and can accept at least one thousand variables for this optimization problem. On the other hand, since the value of $T_{max}$ has an effect on the number of variables, we could reduce the range of $t$, which means find a feasible $T_{max}$, to reduce the number of variables for solving more complex cases by the hybrid solver. In our system model, the range of $t$ is the working time of the NFV system controller, which means that $T_{max}$ is sufficiently large, and the controller has the freedom to determine how to schedule VNFs. However, the objective of our model is to minimize delays of service chains, and we don't need to provide such a long time tolerance for the solver to schedule VNFs. If we set a big value to $T_{max}$, it will bring a lot of variables to our model, and then we need much more qubits to help solve our QUBO model. Therefore, we leverage a greedy algorithm to assist us in finding a reasonable $T_{max}$. For the proposed greedy algorithm, we rearrange all VNFs in service chains to a service chain, and every function $f_{ij}^{k}$ will be allocated to VM $m$, which processes this function for the shortest time. We set the total processing delay by this greedy algorithm to $T_{max}$, and then the QUBO model is embedded in the quantum annealing hardware by an algorithm. The penalty coefficients of the QUBO model play an important role for the hybrid solver in searching for optimal solutions. Since the penalty coefficient needs to be sufficiently large compared with other values in the QUBO model, before setting up the penalty, we evaluate the maximum value that the objective function can reach. In all case studies, the penalty coefficients are set to about 100 times the value of the maximum objective value. After the penalty pre-processing, the penalty could be fine-tuned according to the output, and so we can reach a more suitable penalty set.  Finally, we leverage a D-Wave hybrid solver through the proposed algorithm to solve the QUBO model. The whole proposed algorithm is presented in Algorithm \ref{algorithm1}.

\begin{algorithm}[htb]  
  \caption{}  
  \label{algorithm1}  
  \begin{algorithmic}[1]
   \REQUIRE 
      parameters, $I$, $J$, $M$;  
      the functions in service chain $i$,  $f_{ij}^{k}$;  
      the set of VMs which can process $f_{ij}^{k}$, $V_{ij}^{k}$;
      the NFV network;
    \ENSURE 
       $\tilde{s}$, $x_{ijm}$, $y_{ijmt}$, $z_{ijmt}$, $p_{ijmt}$;  
    \STATE Set the value of $T_{max}$: run the single person greedy algorithm to get a feasible $T_{max}$;  
    \STATE Set the value of penalty coefficients; 
    \STATE Eqs. \cref{eq15,eq16,eq17,eq18,eq19,eq20,eq21,eq22,eq23,eq24,eq25,eq26}: transform from \cref{eq1,eq2,eq3,eq4,eq5,eq6,eq7,eq8,eq9,eq10,eq11,eq12,eq13};
    \STATE The QUBO model: add all terms in \cref{eq15,eq16,eq17,eq18,eq19,eq20,eq21,eq22,eq23,eq24,eq25,eq26} to the right hand side of \eqref{eq15};
    \STATE Embedding the QUBO model onto the quantum annleaing hardware;
    \RETURN $\tilde{s}$, $x_{ijm}$, $y_{ijmt}$, $z_{ijmt}$, $p_{ijmt}$; 
  \end{algorithmic}  
\end{algorithm} 
\section{Experiment}

\begin{table*}[htbp]
\renewcommand\arraystretch{1.4}
\caption{Simulation Results}

\label{table3}
\begin{center}

\begin{tabular}{|c|c|c|c|c|c|c|}
\hline
\textbf{Case} &\textbf{Parameters}&\multirow{2}{3cm}{\centering \textbf{Result of the} \\\textbf{ greedy algorithm $(s)$}} & \multirow{2}{1.5cm}{\centering \textbf{Objective} \\\textbf{Solution $(s)$}} & \multirow{2}{1.8cm}{\centering \textbf{The Longest} \\\textbf{Delay $(s)$}}& \multirow{2}{3.2cm}{\centering \textbf{Average Processing Time} \\ \textbf{for Each VM $(s)$}} & \textbf{Matrx $Q$ Size}\\
&&&&&&\\
\hline
$1$& $I = 2, J = 2, M = 2$& $7$ & $8$ & $5$ & $4.0$ & $(280, 280)$ \\
\hline
$2$& $I = 2, J = 2, M = 2$& $9$ & $11$ & $6$ & $5.5$ & $(338, 338)$ \\
\hline
$3$& $I = 2, J = 2, M = 2$& $5$ & $6$ & $4$ & $3.0$ & $(200, 200)$ \\
\hline
$4$& $I = 2, J = 3, M = 2$& $12$ & $16$ & $9$ & $8.0$ & $(662, 662)$ \\
\hline
$5$& $I = 2, J = 3, M = 2$& $12$ & $20$ & $11$ & $10.0$ & $(662, 662)$ \\
\hline
$6$& $I = 2, J = 3, M = 2$& $19$ & $36$ & $19$ & $18.0$ & $(1012, 1012)$ \\
\hline
$7$& $I = 3, J = 2, M = 2$& $14$ & $17$ & $11$ & $8.5$ & $(732, 732)$ \\
\hline
$8$& $I = 3, J = 2, M = 2$& $10$ & $16$ & $9$ & $8.0$ & $(540, 540)$ \\
\hline
$9$& $I = 3, J = 2, M = 2$& $13$ & $22$ & $12$ & $11.0$ & $(684, 684)$ \\
\hline
$10$& $I = 3, J = 3, M = 2$& $15$ & $49$ & $18$ & $24.5$ & $(1462, 1462)$ \\
\hline
$11$& $I = 3, J = 3, M = 2$& $13$ & $29$ & $13$ & $14.5$ & $(1173, 1173)$ \\
\hline
$12$& $I = 3, J = 3, M = 2$& $14$ & $39$ & $15$ & $19.5$ & $(1266, 1266)$ \\
\hline

\end{tabular}
\end{center}
\noindent \vspace{-0.4cm}
\end{table*}

We study the cases with different parameters $I$, $J$, and $M$, and different service chains, and analyze the performance of the quantum annealer under different cases. $\Delta T$ is equal to $1s$ in all cases. If the hybrid solver cannot output a solution, we will increase the value of $T_{max}$ until the hybrid solver can provide a feasible solution. All results are listed in Table \ref{table3}. In Table \ref{table3}, the third column shows the results of the proposed greedy algorithm, which are feasible solutions of our model and can be used to evaluate the outputs of the D-Wave hybrid solver. The fourth column is the solutions given by the D-Wave hybrid solver, which is the total processing delays of all service chains. The fifth column presents the processing delays of the most time-consuming service chain in each case given by the D-Wave hybrid solver. The sixth column shows the average processing time of each VM, which impacts the costs of resources. The longer the average processing time, the more cost of electricity. The seventh column is the sizes of matrix $Q$ denoted in the QUBO formulation definition in \eqref{eq14}. As the matrix $Q$ size increases, the solver needs to employ more qubits, and the difficulty of solving the problem increases. In Table \ref{table3}, we can find that for case $10$ and case $12$, the longest delay given by the D-Wave hybrid solver is longer than the delay given by the proposed greedy algorithm. It means that the solution given by the hybrid solver cannot be the optimal solution, which shows the hybrid solver cannot solve our model on such a large scale.  

\begin{figure}[htbp]
\centerline{\includegraphics[scale=0.35]{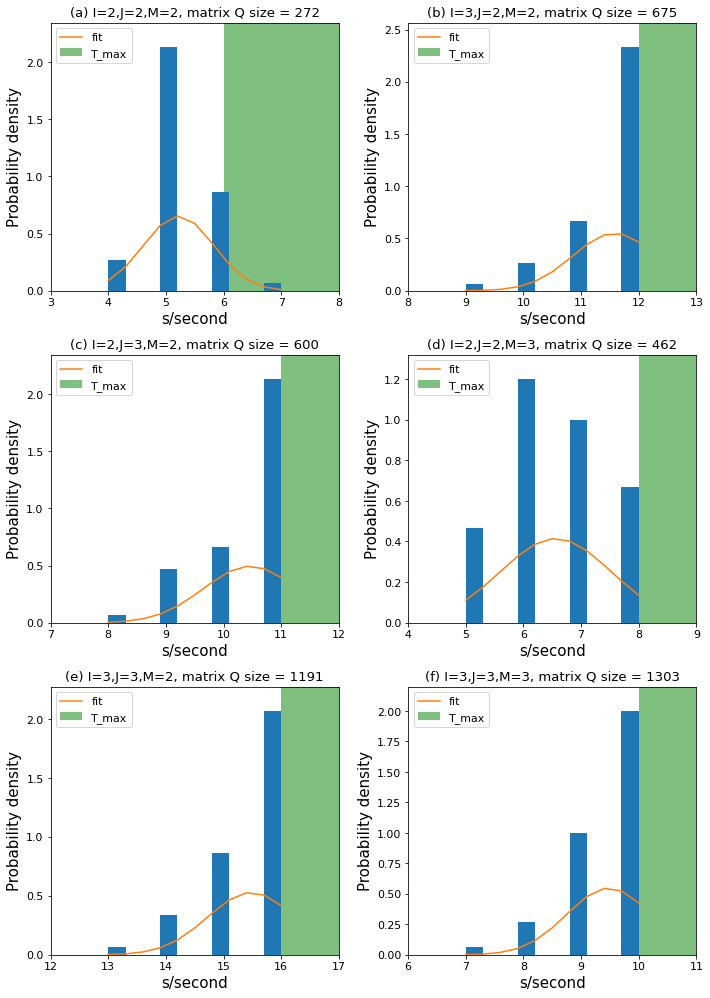}}
\caption{Histograms of the most time-consuming service chain processing delays given by the D-Wave hybrid solver $(a)$ case $a$ $(b)$ case $b$ $(c)$ case $c$ $(d)$ case $d$ $(e)$ case $e$ $(f)$ case $f$.}
\label{fig2}
\noindent \vspace{-0.5cm}
\end{figure}

\begin{figure}[htbp]
\centerline{\includegraphics[scale=0.35]{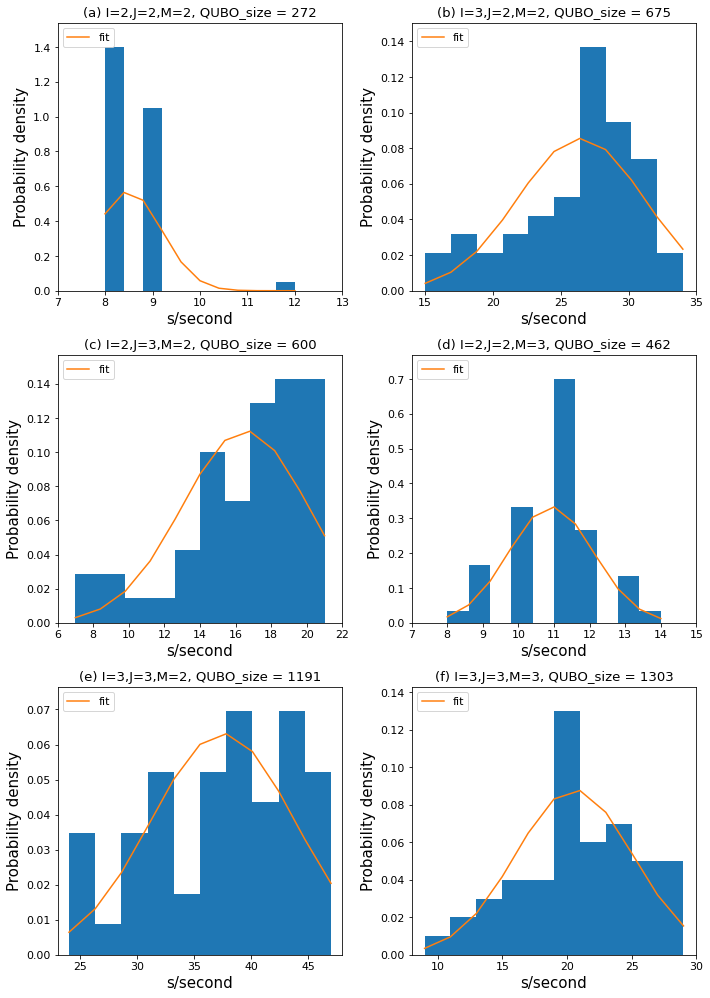}}
\caption{Histograms of the total processing delays of all service chains given by the D-Wave hybrid solver $(a)$ case $a$ $(b)$ case $b$ $(c)$ case $c$ $(d)$ case $d$ $(e)$ case $e$ $(f)$ case $f$.}
\label{fig3}
\noindent \vspace{-0.8cm}
\end{figure}

Fig. \ref{fig2} and Fig. \ref{fig3} are the probability distribution of the results for running 50 times for each case. For all cases, $\Delta T$ is equal to $1s$. Fig. \ref{fig2} shows the most time-consuming service chain processing delays, and Fig. \ref{fig3} shows the total processing delays of all service chains.  In Fig. \ref{fig2},  we can find a higher probability of achieving optimal solutions when the matrix $Q$ size is small. With the increase of matrix $Q$ size, the highest probability of the longest delay moves to $T_{max}$. It means that the difficulty of finding the optimal solution increases as the matrix $Q$ size increases. In Fig. \ref{fig3}, we can find that for case $a$, it is effortless to find the optimal solution by the hybrid solver. For the case $a$ and case $b$, the sizes of matrix $Q$ are small, and solutions are concentrated. For other cases, the sizes of matrix $Q$ are larger, and solutions are more dispersed. Table \ref{table4} shows the hybrid solver running time and the QPU working time for each case. The hybrid solver spends a much longer time on finding a feasible solution for case $f$. Unfortunately, the solver only has the success rate of 4\% to solve case $f$, which means if you leverage the hybrid solver to study case $f$ 100 times, you can only get a feasible solution 4 times. In other words, it is difficult for the hybrid solver to solve this case due to such a large matrix $Q$ size, which corresponds to the results in Fig. \ref{fig2} and Fig. \ref{fig3}. 

\noindent\begin{center}
\begin{table}[htbp]
\renewcommand\arraystretch{1.3}
\caption{\centering{Time Consuming and Success Rate}}
\label{table4}
\begin{tabular}{|c|c|c|c|}
\hline
\textbf{Case} &\multirow{2}{3cm}{\centering \textbf{Average QOU} \\\textbf{access time $(s)$}} & \multirow{2}{2cm}{\centering \textbf{Average solver} \\\textbf{run time $(s)$}} & \textbf{Sucess rate}\\
&&&\\
\hline
$a$& $0.065$ & $2.993$ & $100\%$  \\
\hline
$b$& $0.065$ & $2.997$ & $ 64\% $ \\
\hline
$c$& $0.063$& $2.998$ & $36\%$  \\
\hline
$d$& $0.061$& $2.994$ & $100\%$ \\
\hline
$e$& $0.064$& $2.997$ & $58\%$ \\
\hline
$f$& $0.063$& $3.630$ & $4\%$ \\
\hline
\end{tabular}
\noindent \vspace{-0.3cm}
\end{table}
\par\end{center}

\section{Conclusion}

In this paper, we formulate the VNFs scheduling problem as an ILP model with the optimization of delay and transfer the ILP model into the QUBO form, which can be solved by the quantum annealing machine. We propose a heuristic algorithm to solve the QUBO formulation, using the D-wave hybrid solver. We report and analyze the solutions of several cases under different settings. Our work shows the possibility of using a quantum computer to allocate resources in NFV. From the results of the case study, we can find that the performance of the hybrid solver is better in cases with fewer variables. There is less probability to achieve the optimal solution in cases with more variables.

\bibliographystyle{IEEEtran}
\bibliography{IEEEabrv,IEEEexample,ref-NFV-QC}

\begin{appendices}
\section{ }
\label{appendices}
For the QUBO formulation, there is only an objective function. Therefore, all constraints in the ILP model must be reformulated into quadratic penalties, and then we add them to the original objective function to form the QUBO formulation. All penalties and their explanations are listed below. In all penalty terms, all penalty coefficients, denoted as $P$ with subscripts, are sufficiently large positive constants. 

We combine \eqref{eq1} and \eqref{eq2}, and have

\noindent \vspace{-0.8cm}
\noindent\begin{center}
\begin{equation}
\min_{p_{iJmt}}\quad\tilde{s}=\sum\limits^{I}_{i=1}\sum\limits_{m=1}^{M}\sum\limits_{t=1}^{T_{max}} p_{iJmt}\cdot (t-1)\cdot \Delta T. \label{eq15}
\end{equation}
\par\end{center}

\noindent \vspace{-0.8cm}
\noindent\begin{center}
\begin{equation}
P_{1ij}\left(\sum\limits_{m\in V^{k}_{ij}}x_{ijm} - 1 \right)^2,\quad \forall i,j.\label{eq16}
\end{equation}
\par\end{center}

\noindent \vspace{-0.8cm}
\noindent\begin{center}
\begin{equation}
P_{1ijm}\left(\sum\limits_{t=1}^{T_{max}}z_{ijmt} - x_{ijm} \right)^2,\quad \forall i,j,m.\label{eq17}
\end{equation}
\par\end{center}
Eq. \eqref{eq16} is transformed from \eqref{eq3}. We find that constraint \eqref{eq3} only allows one $x_{ijm}$, which $m \in V_{ij}^{k}$, is equal to $1$. In \eqref{eq16}, if more than one $x_{ijm}$ is equal to $1$, these terms will add a huge value to the objective function. Thus, the optimizer will avoid this situation. This is why constraints \eqref{eq3} can be transformed as \eqref{eq16}. Eq. \eqref{eq17} is transformed from \eqref{eq4}. Constraint \eqref{eq4} only allows the situation that $x_{ijm} = 1$ and $\sum\nolimits_{t=1}^{T_{max}}z_{ijmt} = 1$ and the situation that $x_{ijm} = 0$ and $\sum\nolimits_{t=1}^{T_{max}}z_{ijmt} = 0$. Eq. \eqref{eq17} has the same effect on the model. Therefore, we transform constraint \eqref{eq4} into the terms in \eqref{eq17}. We transform \eqref{eq5} to the following equation.

\noindent \vspace{-0.8cm}
\noindent\begin{center}
\begin{equation}
P_{mt}\left(\sum\limits_{i\neq i'or j\neq j'}\left(y_{ijmt}\cdot y_{i'j'mt}\right)\right),\quad\forall m,t.\label{eq18}
\end{equation}
\par\end{center}
Constraint \eqref{eq5} shows that either or neither $y_{ijmt}$ can be equal to $1$. If any two $y_{ijmt}$ are equal to $1$ in \eqref{eq18}, these terms will lead the solution away from the minimum. Therefore, constraint \eqref{eq5} is equivalent to the terms in \eqref{eq18}.
We transform \eqref{eq6} to the following equation.
\noindent \vspace{-0.8cm}
\noindent\begin{center}
\begin{equation}
\begin{aligned}
P_{1ijmt}\Bigg( y_{ijmt} - x_{ijm} + r_{1ijmt} \Bigg)^{2}, \quad \forall i,j,m,t.\label{eq19}
\end{aligned}
\end{equation}
\par\end{center}
We need to add slack variables to convert the inequalities in \eqref{eq6} into equalities. We only add one binary slack variable to constraint \eqref{eq6} because the difference between the right hand side and the left hand side must be equal to or less than $1$. If $y_{ijmt} = 1$ and $x_{ijm} = 0$, these terms will add a huge value to the objective function. Thus, the optimizer will avoid this situation. This is why constraint \eqref{eq6} can be transformed as the terms in \eqref{eq19}. $r_{1ijmt}$ is a binary slack variable. We transform \eqref{eq7} to the following equation.

\noindent \vspace{-0.4cm}
\noindent\begin{center}
\begin{equation}
P_{2ijm}\left(\sum\limits_{t=1}^{T_{max}} y_{ijmt} - T_{ijm} \right)^2, \quad\forall i,j;\quad m\in V^{k}_{ij}.\label{eq20}
\end{equation}
\par\end{center}
Constraint \eqref{eq7} ensures that the number of $y_{ijmt}$ valued $1$ is $T_{ijm}$. If the number of $y_{ijmt}$ valued $1$ is not equal to $T_{ijm}$, the terms in \eqref{eq20} will add a large value to the objective function. Thus, the optimizer will avoid this situation. This is why constraint \eqref{eq7} can be transformed as the terms in \eqref{eq20}. We transform \eqref{eq8} to the following equation.

\noindent \vspace{-0.8cm}
\noindent\begin{center}
\begin{equation}
P_{2ijmt}\Big(z_{ijmt}\cdot p_{ijmt}\Big),\quad\forall i,j,m,t.\label{eq21}
\end{equation}
\par\end{center}
Constraint \eqref{eq8} precludes the situation that both $z_{ijmt}$ and $p_{ijmt}$ are equal to $1$. All terms in \eqref{eq21} have the same effect as constraint \eqref{eq8}, and so we transform \eqref{eq8} as the terms in \eqref{eq21}. We transform \eqref{eq9} to the following equation.

\noindent \vspace{-0.8cm}
\noindent\begin{center}
\begin{equation}
\begin{aligned}
P_{3ijmt}\Big( y_{ijm(t-1)} - y_{ijmt} + z_{ijmt} - p_{ijmt} \Big)^{2}, \\
\forall i,j,m,t.\quad\qquad\qquad\qquad\label{eq22}
\end{aligned}
\end{equation}
\par\end{center}

\noindent \vspace{-0.8cm}
\noindent\begin{center}
\begin{equation}
\begin{split}
P_{4ijmt} & \left(\sum\limits_{\alpha=1}^{T_{ijm}} z_{ijm(t-\alpha+1)} - y_{ijmt} + r_{2ijmt} \right)^2 , \\ 
& \qquad\qquad\forall i,j,t;\quad m\in V^{k}_{ij}.\label{eq23}
\end{split}
\end{equation}

\par\end{center}

\noindent \vspace{-0.8cm}
\noindent\begin{center}
\begin{equation}
\begin{aligned}
P_{1ijm't} \Bigg( {z_{i(j+1)m't} - \sum\limits_{m\in V_{ij}^{k}}\sum\limits_{\beta=1}^{T_{max}} p_{ijm(t-\beta+1)}} \\
{+ r_{ijm't}} \Bigg)^{2},\quad\forall i,j,t; \quad m'\in V^{k'}_{i(j+1)}.\label{eq24}
\end{aligned}
\end{equation}
\par\end{center}
Eq. \eqref{eq23} is equivalent to \eqref{eq10}. $r_{2ijmt}$ is a binary slack variable.
Eq. \eqref{eq24} is transformed from \eqref{eq11}. We need to add slack variables to convert the inequalities in \eqref{eq11} into equalities. We only add one binary slack variable to constraint \eqref{eq11} because the difference between the right hand side and the left hand side must be equal to or less than $1$. $r_{ijm't}$ is a binary slack variable. We transform \eqref{eq12} and \eqref{eq13} to the following equations.

\noindent\begin{center}
\noindent \vspace{-0.5cm}
\begin{equation}
\begin{split}
 & P_{3ijm} \cdot x_{ijm}^{2} +P_{5ijmt} \cdot y_{ijmt}^{2} + P_{6ijmt} \cdot z_{ijmt}^{2}\\
  + & P_{7ijmt} \cdot p_{ijmt}^{2},\quad\forall i,j,t; \quad m\notin V_{ij}^{k}. \label{eq25}
 \end{split}
\end{equation}

\begin{equation}
\begin{split}
& P_{3ijm}\Bigg(\sum\limits_{m\in V_{ij}^{k}}\sum\limits_{t=1}^{T_{max}}z_{ijmt} - 1 \Bigg)^{2}\\
+ & P_{4ijm}\Bigg( \sum\limits_{m\in V_{ij}^{k}}\sum\limits_{t=1}^{T_{max}} p_{ijmt}- 1 \Bigg)^{2},\quad\forall i,j. \label{eq26}
 \end{split}
\end{equation}

\par\end{center}
To form the QUBO formulation, all terms in \cref{eq16,eq17,eq18,eq19,eq20,eq21,eq22,eq23,eq24,eq25,eq26} need to be added to the right hand side of \eqref{eq15}. 

\end{appendices}
  
\end{document}